\documentclass[a4paper,12pt]{article}

\title{International Stock Market Efficiency:\\
A Non-Bayesian Time-Varying Model Approach\thanks{We would like to thank the editor, Mark Taylor, two anonymous referees, Kenjiro Hirayama, Jun Ma, Daisuke Nagakura, Toshiaki Watanabe, Wako Watanabe, Tomoyoshi Yabu, Yohei Yamamoto, Makoto Yano, seminar participants at Chukyo University Institute of Economics, Doshisha University, conference participants at the Japanese Economics Association 2012 Autumn Meeting in Kyushu Sangyo University, the 10th Biennial Pacific Rim Conference of the Western Economic Association International in Tokyo, the 88th Annual Conference of the Western Economic Association International in Seattle, and the Japanese Economics Association 2013 Autumn Meeting in Kanagawa University for their helpful comments and suggestions. We would also like to thank the financial assistance provided by the Japan Society for the Promotion of Science Grant in Aid for Scientific Research No.24530364 and the Zengin Foundation for Studies on Economics and Finance. All data and programs used for this paper are available upon request.}}%
\author{\textbf{Mikio Ito}\thanks{Faculty of Economics, Keio University, 2-15-45 Mita, Minato-ku, Tokyo 108-8345, Japan (ito@econ.keio.ac.jp, Tel. +81-3-5427-1284).}\\
\bigskip
\textbf{\normalsize Keio University} \and \textbf{Akihiko Noda}\thanks{Corresponding Author: Faculty of Economics, Wakayama University, 930 Sakaedani, Wakayama 640-8510, Japan (noda@eco.wakayama-u.ac.jp, Tel. +81-73-457-7712).}\\
\bigskip
\textbf{\normalsize Wakayama University} \and \textbf{Tatsuma Wada}\thanks{Department of Economics, Wayne State University, Faculty and Administration Building, 656 W.Kirby St., Detroit, MI, 48202 (tatsuma.wada@wayne.edu, Tel. 313-577-3001).}\\
\bigskip
\textbf{\normalsize Wayne State University}}
\date{\empty}
\usepackage[dvips]{graphicx}

\usepackage[]{natbib}%
\usepackage{amsmath,amssymb}%
\usepackage{ascmac}%
\usepackage{multirow}%
\usepackage{lscape}%
\usepackage{subfigmat}

\usepackage{pifont}%
\usepackage{arydshln}%
\usepackage[format=hang]{caption}
\usepackage[all]{xy}
\usepackage{url}
\bibpunct{(}{)}{;}{a}{}{,}

\def\hsymbu#1{\smash{\lower1.7ex\hbox{\huge$#1$}}}

\def\ve #1{{\mbox{\boldmath $#1$}}}

\newcommand{\citetapos}[1]{\citeauthor{#1}'s \citeyearpar{#1}}

\def\ve #1{{\mbox{\boldmath $#1$}}}

\begin{document}

\maketitle %
\bibliographystyle{asa}%

\begin{abstract}
 This paper develops a non-Bayesian methodology to analyze the time-varying structure of international linkages and market efficiency in G7 countries. We consider a non-Bayesian time-varying vector autoregressive (TV-VAR) model, and apply it to estimate the joint degree of market efficiency in the sense of \citet{fama1970ecm,fama1991ecm}. Our empirical results provide a new perspective that the international linkages and market efficiency change over time and that their behaviors correspond well to historical events of the international financial system.\\

\\
\end{abstract}

\pagebreak

\section{Introduction}\label{sec:inter_market_intro}

How are the world's stock markets linked to each other? Is it possible that a boom in one country's stock market could be accompanied by a boom in another country's stock market, while a drop in equity prices occurs simultaneously in many countries' stock markets?  Recent technological progress in the financial sector has enabled information and funds to be rapidly transmitted, thereby providing investors with many opportunities in world stock markets, rather than in just a local stock market. Also, economic activities of firms have become more international, mostly thanks to low transportation costs stemming from technological progress. The answer to the aforementioned (second) question seems to be affirmative in such an environment. 

If so, what are implications of the international linkage for \citetapos{fama1970ecm} market efficiency, which requires zero unexploited excess profit given the information available to the public? For example, consider the case where one nation's stock market is not efficient, but that stock market is jointly efficient with another country's stock market. One possible interpretation for this phenomenon would be that investors have opportunities to invest in two markets -- as a result of portfolio diversification -- and arbitrages occur across these two markets. Hence, it is conceivable that the ``joint efficiency'' among several markets appears when those markets are highly integrated.\footnote{As described in Section 2, the idea of joint efficiency is not new; for example, \citet{macdonald1989fem} focus on the relationship between cointegration and joint efficiency in foreign exchange markets.}

As we shall see in the next section, which reviews the literature, there are mainly two types of papers written in the 1990s and early 2000s.

The first approach employs a vector autoregressive (VAR) model to determine whether there is any international linkage of stock prices, especially in short-run relationships among stock price indices (see, for example, \citet{jeon1990gic} and \citet{tsutsui2004als}). Studies in the second category, on the other hand, shed more light on the long-run equilibrium relationship among returns on stocks by using a vector error correction (VEC) model and cointegration tests (see, for example, \citet{kasa1992cst}, \citet{chan1997ism}). Further, some of the studies examine short-run deviations of returns from long-run relationships by evaluating variance decomposition and impulse response functions.

What is common in the two approaches is that they assume constant parameters in their VAR or VEC with few exceptions that consider structural changes (\citet{narayan2006rwm}, for example). In other words, the relationships appearing in the models are characterized by the parameters that are invariant over time. This assumption seems too strong or even implausible for international financial markets where regime switchings or structural changes occur perhaps due to policy changes and the increasing presence of emerging markets. Furthermore, given the fact that technological changes have been connecting global equity markets more and more strongly -- by facilitating instant and international trades --, the assumption of the stable relationships among international markets is not appealing.  

In order to take into account regime shifts or structural changes, traditionally, researchers have often used subsamples of the data to exclude such aberrant events. By splitting the whole sample into several subsamples so that each subsample does not contain a break or shift, one can easily estimate the relationship among variables within the subsamples. The biggest challenge to this method is to find the cutoff points that determine subsamples.  

Another approach that is preferred by some economists is the so-called rolling window method. Instead of using subsamples defined by breaks, this method allows a researcher to slide the fixed-width window (i.e., subsample) by a given increment. For example, to find the correlation coefficient of two variables at time-$t$, one can utilize data from $t-h$ to $t+h$ as a subsample; likewise, the correlation coefficient at $t+1$ needs data from $(t+1)-h$ to $(t+1)+h$. Here, the width of the window is $2h+1$. In effect, to estimate the correlation coefficient for all $t$, one must use overlapping subsamples. The remaining problem for this approach is that a researcher has to determine the width of the window (or $h$).

The proposed approach in this paper is free from choices of cutoff points or the width of windows. In fact, we are able to demonstrate that our method, a non-Bayesian time-varying VAR (TV-VAR) model, can quite easily be implemented. This result is in line with that of \citet{ito2012eme}, who propose the TV-VAR model for a univariate case. Furthermore, by defining and computing the time-varying degree of the market efficiency from the estimated TV-VAR, we are able to discover the time-varying structure of world stock markets. This time-varying degree of market efficiency is used in conjunction with its statistical inference -- which is calculated by our new bootstrap method -- to determine whether or not the stock market is efficient.

To sum up, this paper's contribution is two-fold. First, we look at the linkage of international equity markets through the novel non-Bayesian TV-VAR that possesses statistically more preferable properties than alternatives. This is a clear deviation from previous studies that employ standard VARs and VECs. As the second contribution, we propose the ``time-varying degree of market efficiency'' associated with the TV-VAR model and its statistical inference. This new measure and its statistical inference provide information as to whether the linkage is so strong that investors regard the markets for which efficiency is jointly confirmed as a single market. Our method of evaluating (joint) market efficiency is an alternative to the previous studies that employ the VEC model, where cointegration may imply market inefficiency. This is because (i) we incorporate the time-varying nature of market efficiency and (ii) the evaluation is based on estimated TV-VAR coefficients.

Our empirical results affirm the time-varying structure in the linkages and in the market efficiency of international stock markets. We also find that the time-varying nature of the market structure corresponds well to historical events in the international financial system. 

This paper is organized as follows. Section \ref{sec:inter_market_review} reviews the literature. In Section \ref{sec:inter_market_model}, the model and new methodologies for our non-Bayesian TV-VAR model are presented. We also introduce our new measure, the degree of market efficiency in Section \ref{sec:inter_market_model}. The data on G7 stock markets, together with preliminary unit root test results, are described in Section \ref{sec:inter_market_dat}. In Section \ref{sec:inter_market_er}, our empirical results show that international stock markets are inefficient in the late 1980s; and the European markets become jointly inefficient after the late 1990s. Section \ref{sec:inter_market_cr} concludes. The Online Technical Appendix provides mathematical and statistical discussions about the new estimation methodologies that we have developed.\footnote{Online Technical Appendix is available at \url{http://at-noda.com/appendix/inter_market_appendix.pdf}.}

\section{A Review of the Literature}\label{sec:inter_market_review}

The transmission mechanism of shocks in international equity markets has been widely studied. A natural way to address this question is to apply a VAR to stock price data. For example, \citet{eun1989its} find that a shock originated from the U.S. is transmitted to other markets, while \citet{jeon1990gic} utilize daily stock price data to investigate the correlation structure among stock markets in the world, paying special attention to the 1987 stock market crash. By focusing on Nordic and US stock markets, \citet{mathur1990ian} find that the U.S. market affected the Danish market, but not other Nordic markets. In order to compute the responses of stock markets more precisely, the specification of the VAR, especially the length of lags, has been further investigated by \citet{tsutsui2004als}.
  
Yet, one problem remains. Stock price data are mostly non-stationary. Therefore, there should not be a stable relationship between stock prices in different markets unless they are cointegrated. Naturally, then, researchers' interest has shifted to whether or not stock prices are cointegrated. In addition, as \citet{macdonald1988mpe,macdonald1989fem} point out, if the stock prices of any two markets are cointegrated, then there must be a linear combination of stock prices that helps investors predict the future stock prices; therefore, the markets are not efficient.\footnote{This view, however, is questioned by \citet{dwyer1992cme}.} Subsequent attempts to find cointegrating relationships and associated common trends are made by \citet{jeon1991ssp}, \citet{kasa1992cst}, \citet{corhay1993cst}, \citet{blackman1994ltr}, \citet{chung1994cst}, \citet{choudhry1994sts,choudhry1997sts}, and \citet{chen2002sml}, among others. While \citet{chan1997ism} and \citet{balios2003iid} shed light on market efficiency, by carefully conducting several cointegration tests, \citet{pascual2003aes} calls into question the power of the cointegration test when the sample size increases. With the adjusted power, the cointegration test reveals no long-run relationship among the UK, French and German stock markets.

More recently, studies employ both VAR and VEC, because the international data set almost always includes both stationary and nonstationary series. For example, \citet{cheung2010gcm} assess the spill-over effect of credit risk on the wake of the world financial crisis in 2007-2009; \citet{hirayama2013isp} disentangle the stock price co-movement by considering both common and idiosyncratic shocks.

Finally, deviations from the standard VAR and VEC have recently been intensively studied. Assuming possible structural breaks, \citet{narayan2005csm} examine the cointegrating relationship among New Zealand, Australia and the G7 countries. Non-stable relationships in European stock markets -- mainly as to whether convergence has occurred over time -- are investigated by \citet{pascual2003aes} whose VEC is estimated within rolling samples. Similarly, together with structural break tests and the rolling cointegration analysis, \citet{mylonidis2010des} focus on convergence in the European stock markets.

\section{The Model}\label{sec:inter_market_model}

\subsection{Preliminaries}

First, we consider a simple VAR model to analyze the stock market linkage in the context of market efficiency. Let ${\ve y}_{t}$ denote a $k$-dimensional vector representing the rates of return for $k$ stocks at $t$. More specifically, ${\ve y}_{t}$ consists of $k$ countries' stock market returns. In the literature, a number of previous studies employ VAR($p$) models to analyze the linkage of stock markets:

\begin{equation}
{\ve y}_{t}={\ve\nu}+A_{1}{\ve y}_{t-1}+\cdots +A_{p}{\ve y}_{t-p}+{\ve u}_{t};\text{ \ }t=1,2,\ldots ,T,
\label{VAR}
\end{equation}%
where ${\ve\nu}$ is a vector of intercepts; ${\ve u}_{t}$ is a vector of error terms that are serially uncorrelated. Note that the unexpected, excess returns on stocks at t are solely driven by the error term ${\ve u}_t$ because ${\ve y}_t-E\left[{\ve y}_t \ | \ {\ve y}_{t-1}, {\ve y}_{t-2}, \cdots\right]={\ve u}_t$, where $E\left[{\ve y}_t \ | \ {\ve y}_{t-1}, {\ve y}_{t-2},\cdots\right]$ represents the conditional expectation of the $k$ countries' stock returns at $t$ given the returns at any previous periods $t-1, t-2, \cdots$. This set-up is in accordance with the view of the efficiency market hypothesis (EMH) in the sense that there is no unexploited excess profit given the information available to the public (see \citet{fama1970ecm,fama1991ecm}, for example).

Oftentimes, it is understood that the VAR($p$) model is a reduced form of a data generating process that is a VMA($\infty $) model:%
\begin{eqnarray}
{\ve y}_{t} &=&{\ve\mu} +\Phi _{0}{\ve u}_{t}+\Phi _{1}{\ve u}_{t-1}+\Phi _{2}{\ve u}_{t-2}+\cdots  \notag
\\
&=&{\ve\mu} +\Phi \left( L\right) {\ve u}_{t},  \label{VMA}
\end{eqnarray}%
where $\Phi \left( L\right) $ is a matrix lag polynomial of a lag operator $L$, i.e., $\Phi \left( L\right) =\Phi _{0}+\Phi _{1}L+$ $\Phi_{2}L^{2}+\cdots$, with all eignvalues of $\Phi \left( 1\right)$ being outside the unit circle. One clear assumption here is that coefficient matrices $\left\{ \Phi _{i}\right\} _{i=0}^{\infty }$ are time invariant or parameter matrices of $k\times k$. With such matrices, one can compute the impulse-response functions along with the identification assumptions such as $\Phi _{0}=I$ (an identity matrix). Note that the long-run effect of ${\ve u}_{t}$ on ${\ve y}$ is given by $\Phi \left( 1\right)$; the vector of expected excess returns, $E\left[{\ve y}_t \ | \ {\ve y}_{t-1}, {\ve y}_{t-2},\cdots \right]-{\ve\mu}$, is zero when $\Phi \left( 1\right)=I$.

In this paper, we view equation (\ref{VMA}) as the model and Equation (\ref{VAR}) serves as a device for estimation. Our focus as to whether EMH holds hinges on whether $\Phi \left( 1\right)=I$, or equivalently, whether $\left[ I-A\left( 1\right) \right]^{-1}=\left[I-A_{1}-A_{2}-\cdots-A_{p}\right]^{-1}=I$. The following subsection explains the extended version of our estimation model, equation (\ref{VAR}).

\subsection{Non-Bayesian Time-Varying VAR Model}\label{subsec:tv_var}

When the parameters in our VAR coefficient matrices $\left\{ A_{i}\right\}_{i=0}^{p}$ do not seem to be constant over time, more precisely, when a parameter consistency test such as \citet{hansen1992a} rejects the null hypothesis of constant parameters,\footnote{See Online Technical Appendix A.1.} it is more appropriate to use the time-varying (TV-) VAR model that allows the parameters in $\left\{ A_{i}\right\} _{i=0}^{p}$ to vary with time:
\begin{equation}
{\ve y}_{t}={\ve\nu}+A_{1,t}{\ve y}_{t-1}+\cdots +A_{p,t}{\ve y}_{t-p}+{\ve u}_{t};\text{ \ }t=1,2,\ldots
,T,  \label{TVVAR}
\end{equation}%
with 
\begin{equation}
A_{i,t}=A_{i,t-1}+V_{i,t}  \label{Arw}
\end{equation}%
for $i=1,\ldots ,p$, and $t=1,2,\ldots ,T$. The assumption for the process, Equation (\ref{Arw}), stems from the fact that the alternative hypothesis of \citetapos{hansen1992a} test is the multivariate random walk process. For this time-varying specification, we consider the underlying time-varying VMA model:%
\begin{equation}
{\ve y}_{t}={\ve\mu}_{t}+\Phi _{0,t}{\ve u}_{t}+\Phi _{1,t}{\ve u}_{t-1}+\Phi_{2,t}{\ve u}_{t-2}+\cdots
\label{TVVMA}
\end{equation}%
with $\Phi _{0,t}=I$ for all $t$.

Taken together, the following is the extended version of \citet{ito2012eme}, who consider the time-varying model within a univariate context. In this paper we propose the Non-Bayesian time-varying model with a vector process. First of all, it is convenient for us to set our model in a state-space form, especially for estimation purposes. To this end, the observation Equation (\ref{TVVAR}) is written as%
\begin{equation*}
{\ve y}_{t}={\ve\nu}+A_{t}Z_{t-1}+{\ve u}_{t};\text{ \ }t=1,2,\ldots ,T,
\end{equation*}%
where 
\begin{equation*}
A_{t}=\left[ 
\begin{array}{ccc}
A_{1,t} & \cdots & A_{p,t}%
\end{array}%
\right] \text{ \ and }Z_{t-1}=\left[ 
\begin{array}{ccc}
{\ve y}_{t-1}^{\prime} & \cdots & {\ve y}_{t-p}^{\prime}%
\end{array}%
\right] ^{\prime };
\end{equation*}%
and the state Equations (\ref{Arw}) becomes 
\begin{equation*}
vec\left( A_{t}\right) =vec\left( A_{t-1}\right) +{\ve v}_{t},
\end{equation*}%
where ${\ve v}_{t}=vec\left( V_{t}\right) =vec\left( \left[ 
\begin{array}{ccc}
V_{1,t} & \cdots & V_{p,t}%
\end{array}%
\right] \right) $ is a $k^{2}p$-vector.

Furthermore, to see the advantages of our estimation method, we stack equations (\ref{TVVAR}) from $t=1$ through $T$,
\begin{equation*}
{\ve y}=D{\ve\beta}+{\ve u}
\end{equation*}%
where ${\ve y}=\left( 
\begin{array}{ccc}
{\ve y}_{1}^{\prime} & \cdots & {\ve y}_{T}^{\prime}%
\end{array}%
\right) ^{\prime },$ ${\ve u}=\left( 
\begin{array}{ccc}
{\ve u}_{1}^{\prime } & \cdots & {\ve u}_{T}^{\prime }%
\end{array}%
\right) ^{\prime },$ $D$ is a matrix consisting of $%
\begin{array}{ccc}
Z_{0}, & \cdots & Z_{T-1},%
\end{array}%
$ and $\ve\beta$ is a vector that contains $%
\begin{array}{ccc}
A_{1} & \cdots & A_{T}.%
\end{array}%
$ For more details, see Online Technical Appendix A.2.

The time-varying nature of the VAR coefficients, more specifically Equation (\ref{Arw}), is also stacked from $t=1$ to $T$, and written in a matrix form: 
\begin{equation*}
{\ve\gamma}=W{\ve\beta}+{\ve v}
\end{equation*}%
where a vector $\ve\gamma$ includes $vec\left( A_{0}\right) $ $\ $and zeros; $W$ is a matrix consisting of identity matrices and zeros; and ${\ve v}=\left( 
\begin{array}{ccc}
{\ve v}_{1}^{\prime} & \cdots & {\ve v}_{T}^{\prime}%
\end{array}%
\right) ^{\prime}$. Noticing that both systems of equations share the same vector $\ve\beta$, we combine them together and our entire system now becomes 
\begin{equation}
\left[ 
\begin{array}{c}
{\ve y} \\ 
{\ve \gamma}%
\end{array}%
\right] =\left[ 
\begin{array}{c}
D \\ 
W%
\end{array}%
\right] {\ve\beta} +\left[ 
\begin{array}{c}
{\ve u} \\ 
{\ve v}%
\end{array}%
\right] .  \label{sys1}
\end{equation}

One of the very convenient features of our method is that the system (\ref{sys1}) allows us to estimate the unknown parameters $\beta$ (time-varying coefficients) and the variance-covariance of the error term by OLS.\footnote{See Online Technical Appendix A.3 for more details.} Unlike the traditional method or the Bayesian estimation, the time-varying model does not necessitate the Kalman filtering which requires iterations from $t=1$ to $T$, to find the likelihood value. Indeed, this OLS based method estimates unknown parameters at one time. This method can also handle cases such as heteroscedasticity, serial correlations, and mutual correlations within the vector $\left({\ve u},{\ve v}\right)$ by utilizing the Generalized Least Squares (GLS) method. Note that, when the Kalman filter is applied to this model, mutual correlations in ${\ve u}$ and ${\ve v}$ need special treatments, such as what \citet{anderson1979opt} describe. Other desirable properties of our method are more thoroughly explained in \citet{ito2012eme}.

The next subsection describes a different way to examine whether EMH holds empirically, under our assumption that the structure or autoregressive coefficients on stock returns vary over time. 

\subsection{Time-Varying Degrees of Market Efficiency}\label{subsec:tv_dme}

While our main focus is whether the vector of expected excess returns, $E\left[{\ve y}_t \ | \ {\ve y}_{t-1}, {\ve y}_{t-2},\cdots\right]-\ve\mu$, is zero in Equation (\ref{VMA}), the time-varying VAR model (\ref{TVVAR}) modifies this condition as $E\left[{\ve y}_t \ | \ {\ve y}_{t-1}, {\ve y}_{t-2},\cdots\right]-\ve\mu_{t}=0$ where $\ve\mu_{t}=\left(I-A_{1,t}-\cdots-A_{p,t}\right)^{-1}{\ve u}_t$. Still, with the estimated coefficient matrices obtained by (\ref{sys1}) via OLS (or GLS), we are able to determine if EMH holds by observing the time-varying matrices $A_{1,t},\ldots ,A_{p,t}$ or $\Phi _{1,t},\Phi_{2,t},\ldots $. This is because the TV-VAR model (\ref{TVVAR}) can be seen as a locally stationary model. To implement this strategy, let us consider a matrix
\begin{equation*}
\Phi _{t}\left( 1\right) =\sum_{j=1}^{\infty }\Phi _{j,t},
\end{equation*}%
which is a cumulative sum of the time-varying VMA coefficient matrices which appears in Equation (\ref{TVVMA}). Since this matrix measures the long-run effect of shocks $\{{\ve u}_{\tau}\}_{\tau=t}^{-\infty}$ on stock return $\{{\ve y}_t\}$, we call it the long-run multiplier that is in line with \citet{ito2012eme}. This multiplier, $\Phi_{t}\left( 1\right)$, is, in fact, regarded as a metric measure for market efficiency in the sense that $\Phi_{t}\left( 1\right) =I$ suggests EMH; while its deviations from an identity matrix measure the degree of market (in)efficiency. Once again, we do not impose the identification assumptions about shocks. In other words, we do not label which shock originated from country $j\in 1,\ldots ,k$. Therefore, it is important for us to investigate whether or not $k$ countries' stock markets are all efficient without specifying the roles of the shocks; and hence, measuring the distance between $\Phi _{t}\left( 1\right) $ and $I$ provides us with a good idea as to how $k$ countries' stock markets are closer to or farther from efficiency.

The distance between $\Phi _{t}\left( 1\right) $ and $I$ is measured by the spectral norm: 
\begin{equation}
\zeta _{t}=\sqrt{\max \lambda \left[ \left( \Phi _{t}\left( 1\right)
-I\right) ^{\prime }\left( \Phi _{t}\left( 1\right) -I\right) \right] },
\label{dist}
\end{equation}%
i.e., the square root of the largest eigenvalue of the square matrix $\left(\Phi_{t}\left(1\right)-I\right)^{\prime }\left(\Phi_{t}\left(1\right)-I\right)$ for each $t$. Clearly, the distance becomes zero ($\zeta_{t}=0$) when $\Phi_{t}\left(1\right)=I$. Yet, the distance becomes a large (positive) number as the two matrices deviate from each other, in the sense of the spectral norm. We call $\zeta_{t}$ the degree of market efficiency that measures how close to or far from the efficient markets the actual markets are.\footnote{In case of univariate data, this norm is essentially same as the degree defined by \citet{ito2012eme}.  That is, the degree in this paper is a natural extension of the one in \citet{ito2012eme}.} By computing this for all the sample points, $t$, we are able to know when the markets are efficient (or close enough to indicate that the markets are efficient) and when the markets are obviously inefficient.

From the correspondence between Equations (\ref{TVVAR}) and (\ref{TVVMA}), provided $\left(I-A_{1,t}-\cdots-A_{p,t}\right) $ is non-singular, we have 
\begin{equation*}
\Phi _{t}\left( 1\right) =\left( I-A_{1,t}-\cdots -A_{p,t}\right) ^{-1}.
\end{equation*}%
In practice, we estimate the TV-VAR Equation (\ref{TVVAR}) to obtain $A_{1,t},\ldots,A_{p,t}$, and then, we compute the degree of market efficiency, Equation (\ref{dist}).

\section{Data}\label{sec:inter_market_dat}

We utilize the monthly data on the Morgan Stanley capital index for the
Group of 7 (G7) countries (United States, Canada, United Kingdom, Japan,
Germany, France, and Italy) from December 1969 through March 2013, as
obtained from the Thomson Reuters Datastream. To compute the ex-post stock
return series, we take the first difference of the natural log of the stock
price index.
\begin{center}
(Table \ref{inter_market_table1} around here)
\end{center}
Provided in Table 1, descriptive statistics show that monthly returns on the 
stock market range from Italy's 0.34\% (4.08\% per annum) to the U.K.'s
0.57\% (6.84\% per annum). The average return on the largest stock market,
U.S. is 0.52\% (6.24\% annum) and the third highest after the U.K. and
Canada (0.54\% or\ 6.48\% annum). Volatility for the U.S. is the smallest
among the seven countries, 3.96\%, while that for Italy is 6.17\%, which is
the largest of the seven countries. Except for the U.K. and France, the
largest month-to-month change is negative, not positive, i.e., crashes have
larger magnitudes than booms.

Table \ref{inter_market_table1} also shows the results of a unit root test. 
It is important for our estimation that the variables in the TV-VAR model 
are all stationary. Thus, we examine this by employing the ADF-GLS test of 
\citet{elliott1996eta}. It is then confirmed that for all the variables, the 
ADF-GLS test uniformly rejects the null hypothesis of the variable's having 
a unit root at the 1\% significance level.

\section{Empirical Results}\label{sec:inter_market_er}

\subsection{The Time-Invariant VAR Model}

Having presented the time-varying model, we must verify that the TV-VAR model is more appropriate to use than the traditional, time-invariant VAR model. To do so, we first estimate the time-invariant VAR model, and then apply the parameter constancy test of \citet{hansen1992a} to investigate whether the time-invariant model is a better fit for our data.

Here we consider the parameter stability for five groups using five VAR models. They are: (i) North American markets (U.S. and Canada); (ii) two largest stock markets (U.S. and U.K.); (iii) three largest stock markets (U.S., U.K. and Japan); (iv) European markets (U.K., Germany, France and Italy); and (v) all seven markets.
\begin{center}
(Table \ref{inter_market_table2} around here)
\end{center}
To select the length of the lags, we adopt the Bayesian information criterion (BIC; \citet{schwarz1978edm}). Table \ref{inter_market_table2} presents both estimated coefficients and \citetapos{hansen1992a} joint parameter constancy test statistics that are shown in the last column, under \textquotedblleft $L_{C}.$\textquotedblright\ For all the five models, the parameter constancy test rejects the null hypothesis of constancy at the 1\% level, against the alternative hypothesis stating that the parameter variation follows a random walk process. These results suggest that the time-invariant VAR model does not fit our data; rather, we should use the TV-VAR model for the G7 data.\footnote{Table A.1 in Online Technical Appendix provides the results of a univariate AR process for each country. From ``$L_C$'' statistics, we can confirm that the TV model is more appropriate for all countries.}

\subsection{TV-VAR Model and the Degree of Market Efficiency}

Now, let us focus on the TV-VAR model and the degree of market efficiency associated with the model. Once again, using the spectral norm, we measure the stock markets' deviation from the efficient condition, by Equation (\ref{dist}). For example, considering U.S. and Canadian stock markets, the degree of market efficiency tells us how the two markets are different from the efficient markets. If $\zeta _{t}=0$ for time-$t$, \ the two stock markets are jointly efficient at that time.

\begin{figure}[bp]
 \caption{Time-Varying Degree of Market Efficiency: North America}
 \label{inter_market_fig1}
 \begin{center}
 \includegraphics[scale=0.8]{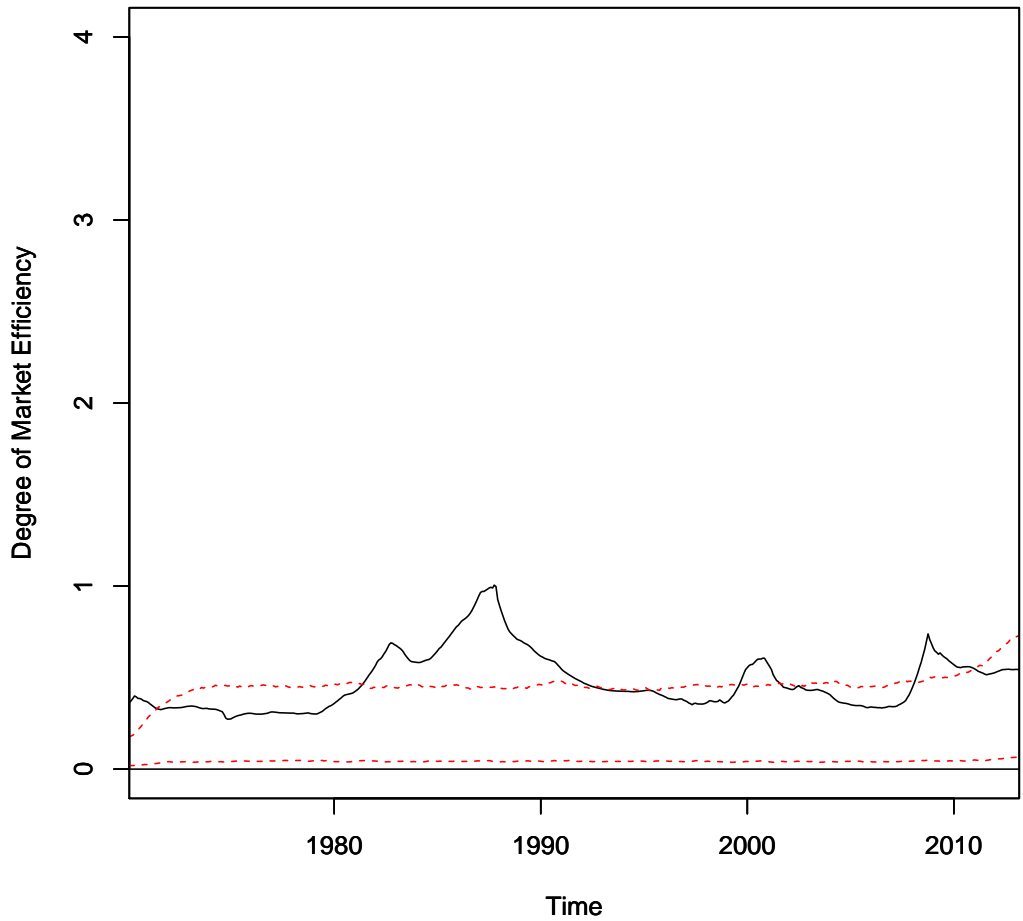}
\vspace*{3pt}
{
\begin{minipage}{350pt}
\footnotesize
\underline{Notes}:
\begin{itemize}
\item[(1)] The dashed red lines represent the 99\% confidence bands of the 
           time-varying spectral norm in case of efficient market. 
\item[(2)] We run 5000 times Monte Carlo sampling to calculate the confidence 
           bands.
\item[(3)] R version 3.1.0 was used to compute the estimates.
\end{itemize}
\end{minipage}}%
\end{center}
\end{figure}

\begin{figure}[bp]
 \caption{Time-Varying Degree of Market Efficiency: U.S. and U.K.}
 \label{inter_market_fig2}
 \begin{center}
 \includegraphics[scale=0.8]{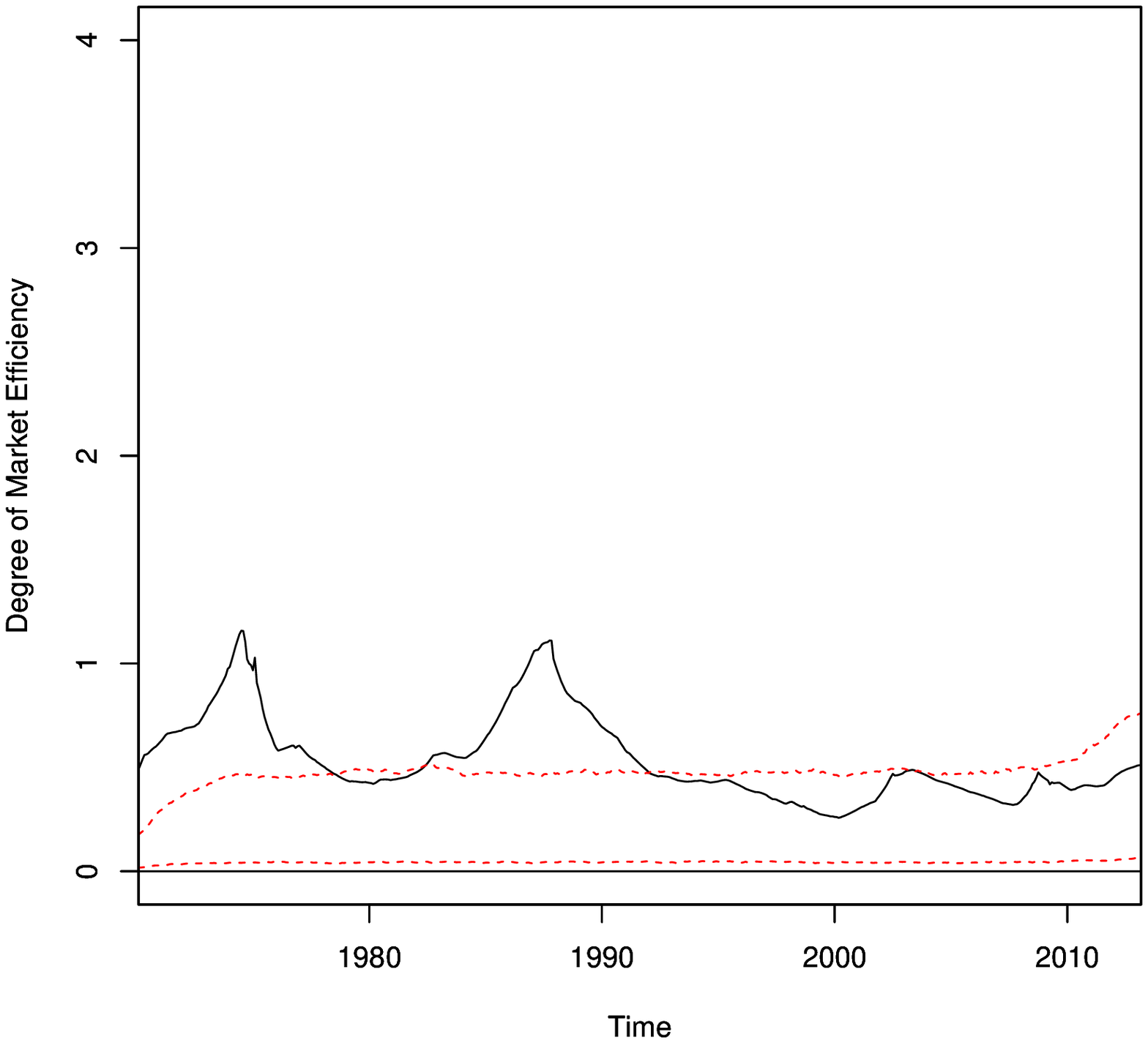}
\vspace*{3pt}
{
\begin{minipage}{350pt}
\footnotesize
\underline{Note}: As for Figure \ref{inter_market_fig1}.
\end{minipage}}%
\end{center}
\end{figure}

\begin{figure}[bp]
 \caption{Time-Varying Degree of Market Efficiency: U.S., U.K. and Japan}
 \label{inter_market_fig3}
 \begin{center}
 \includegraphics[scale=0.8]{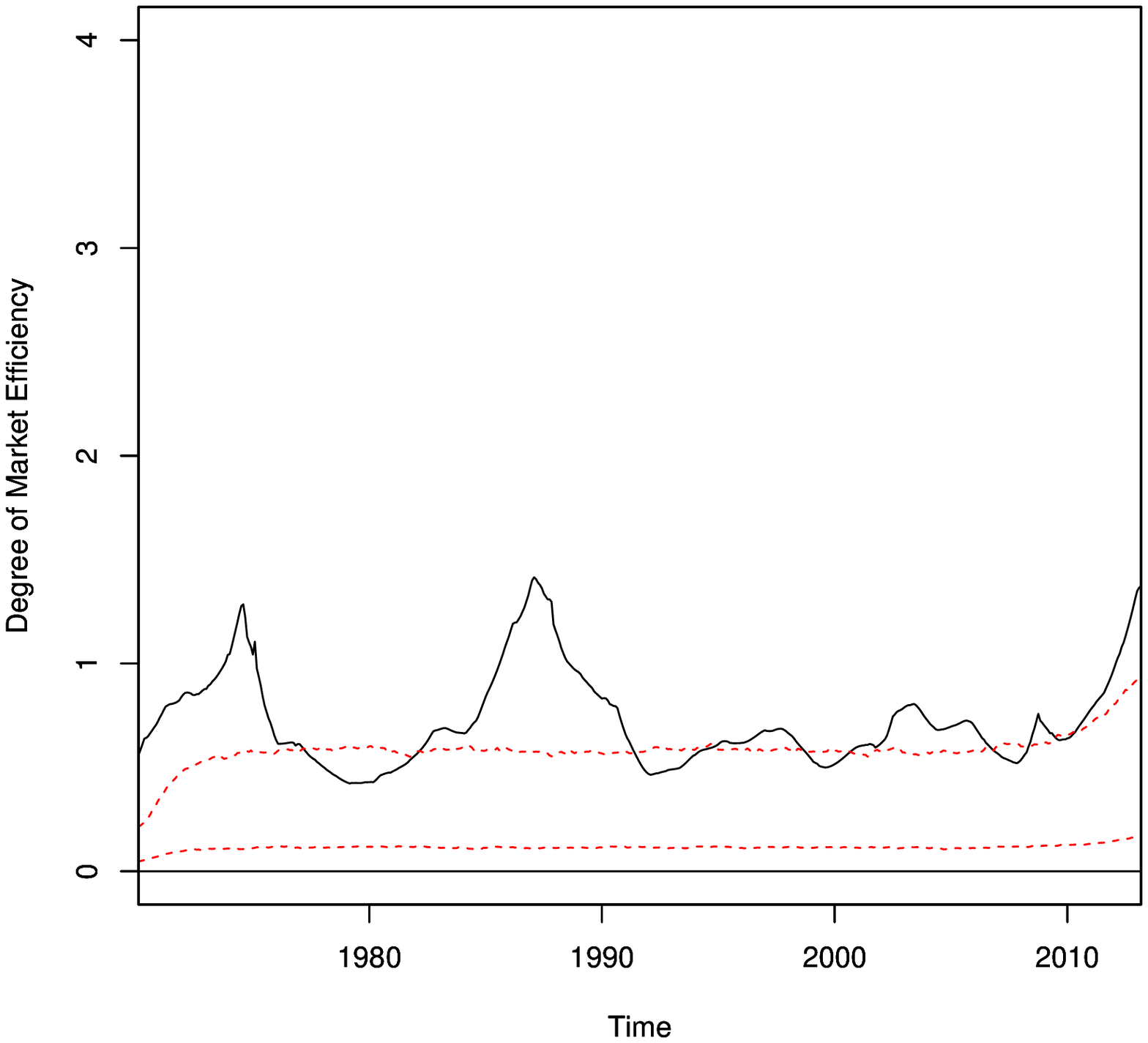}
\vspace*{3pt}
{
\begin{minipage}{350pt}
\footnotesize
\underline{Note}: As for Figure \ref{inter_market_fig1}.
\end{minipage}}%
\end{center}
\end{figure}

\begin{figure}[bp]
 \caption{Time-Varying Degree of Market Efficiency: European Countries}
 \label{inter_market_fig4}
 \begin{center}
 \includegraphics[scale=0.8]{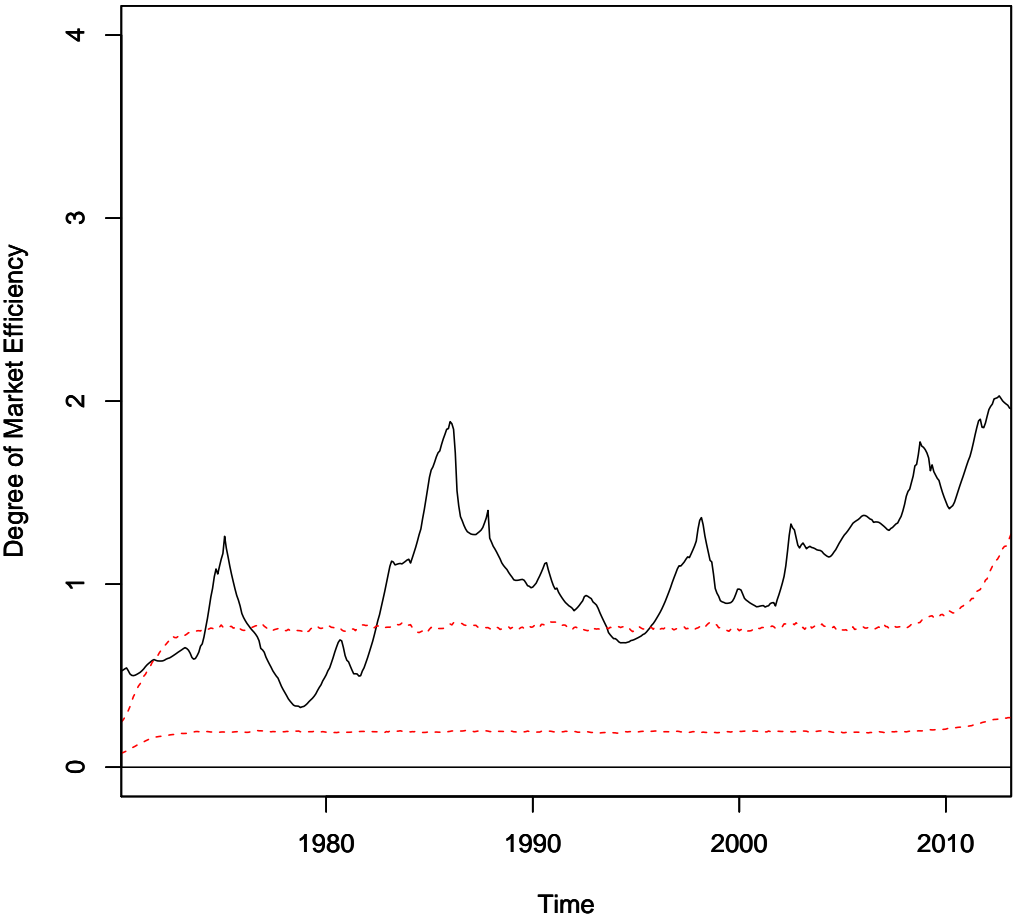}
\vspace*{3pt}
{
\begin{minipage}{350pt}
\footnotesize
\underline{Note}: As for Figure \ref{inter_market_fig1}.
\end{minipage}}%
\end{center}
\end{figure}

\begin{figure}[bp]
 \caption{Time-Varying Degree of Market Efficiency: G7 Countries}
 \label{inter_market_fig5}
 \begin{center}
 \includegraphics[scale=0.8]{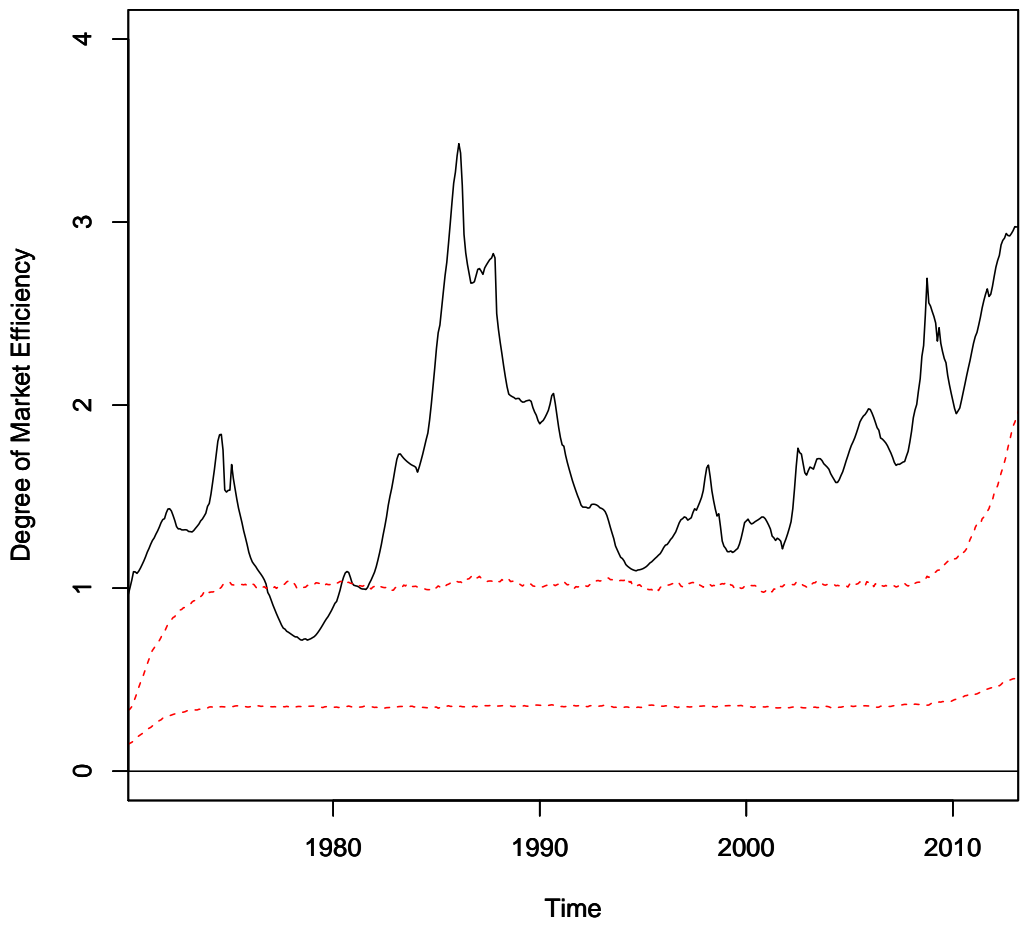}
\vspace*{3pt}
{
\begin{minipage}{350pt}
\footnotesize
\underline{Note}: As for Figure \ref{inter_market_fig1}.
\end{minipage}}%
\end{center}
\end{figure}

\begin{figure}
\caption{Time-Varying Degree of Market Efficiency: Individual Countries}
\label{inter_market_fig6}
\begin{subfigmatrix}{3}
\subfigure[U.S.]{\includegraphics[width=.3\textwidth]{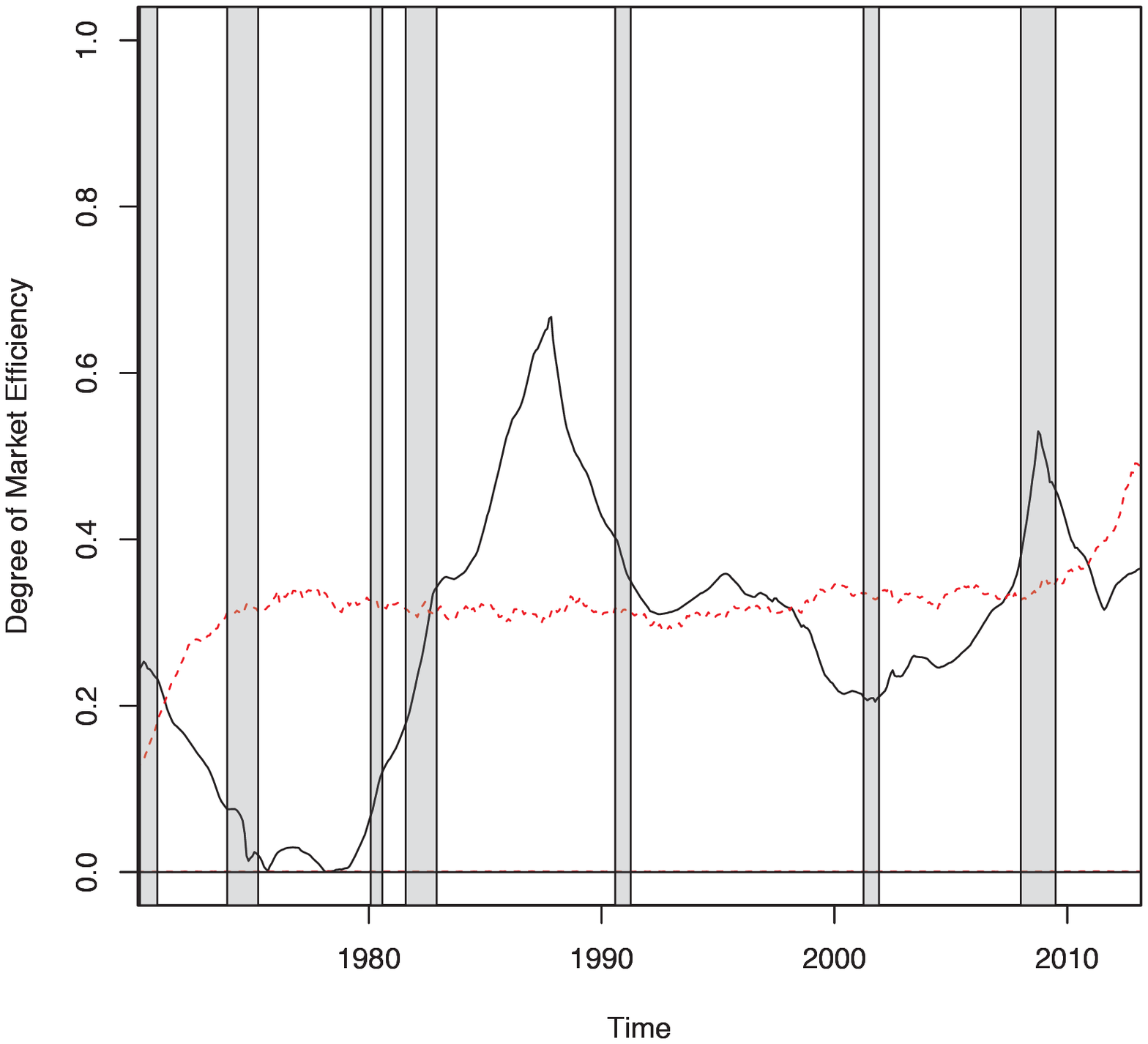}
\label{fig:figb11}}
\subfigure[Canada]{\includegraphics[width=.3\textwidth]{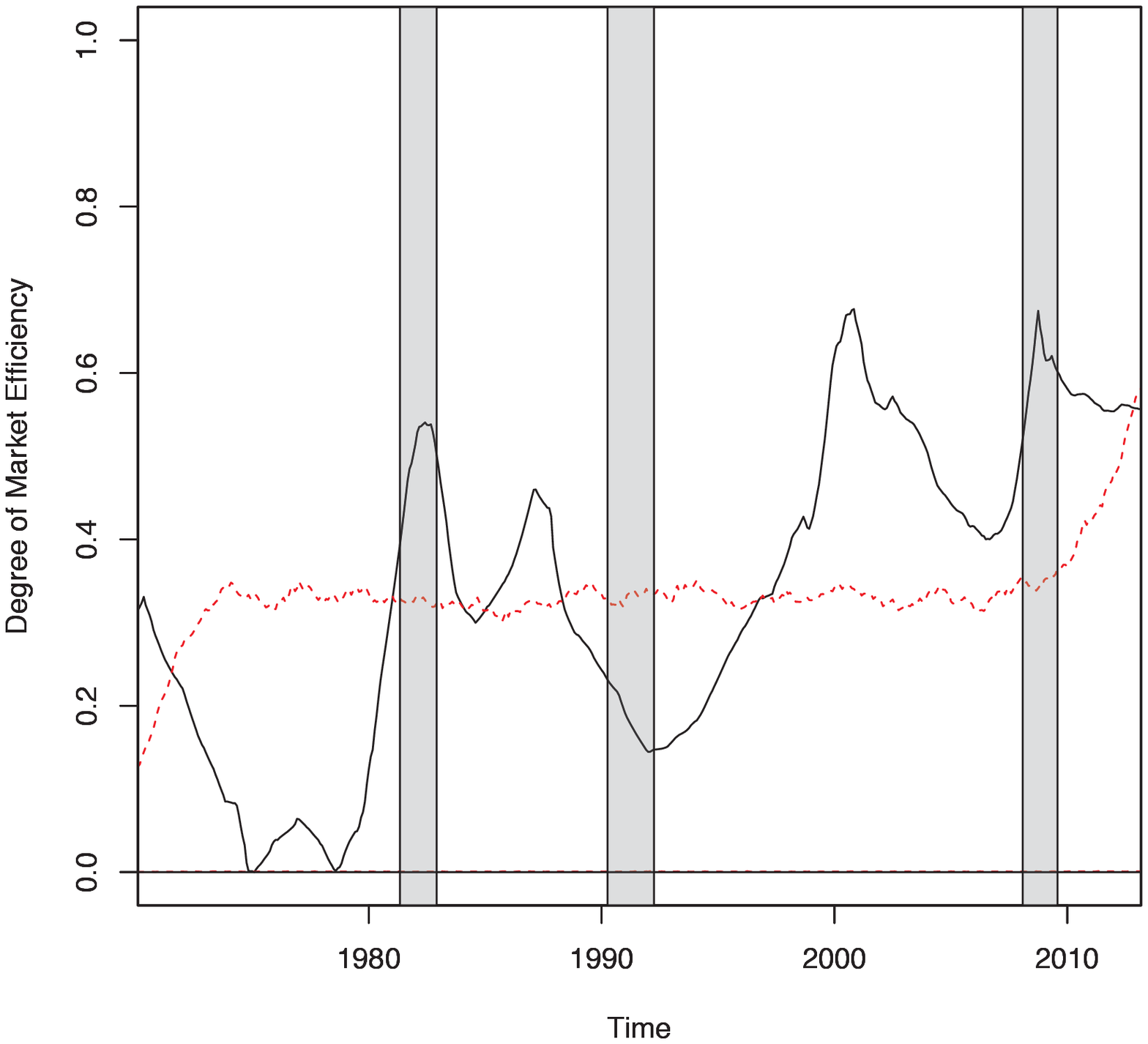}
\label{fig:figb12}}
\subfigure[U.K.]{\includegraphics[width=.3\textwidth]{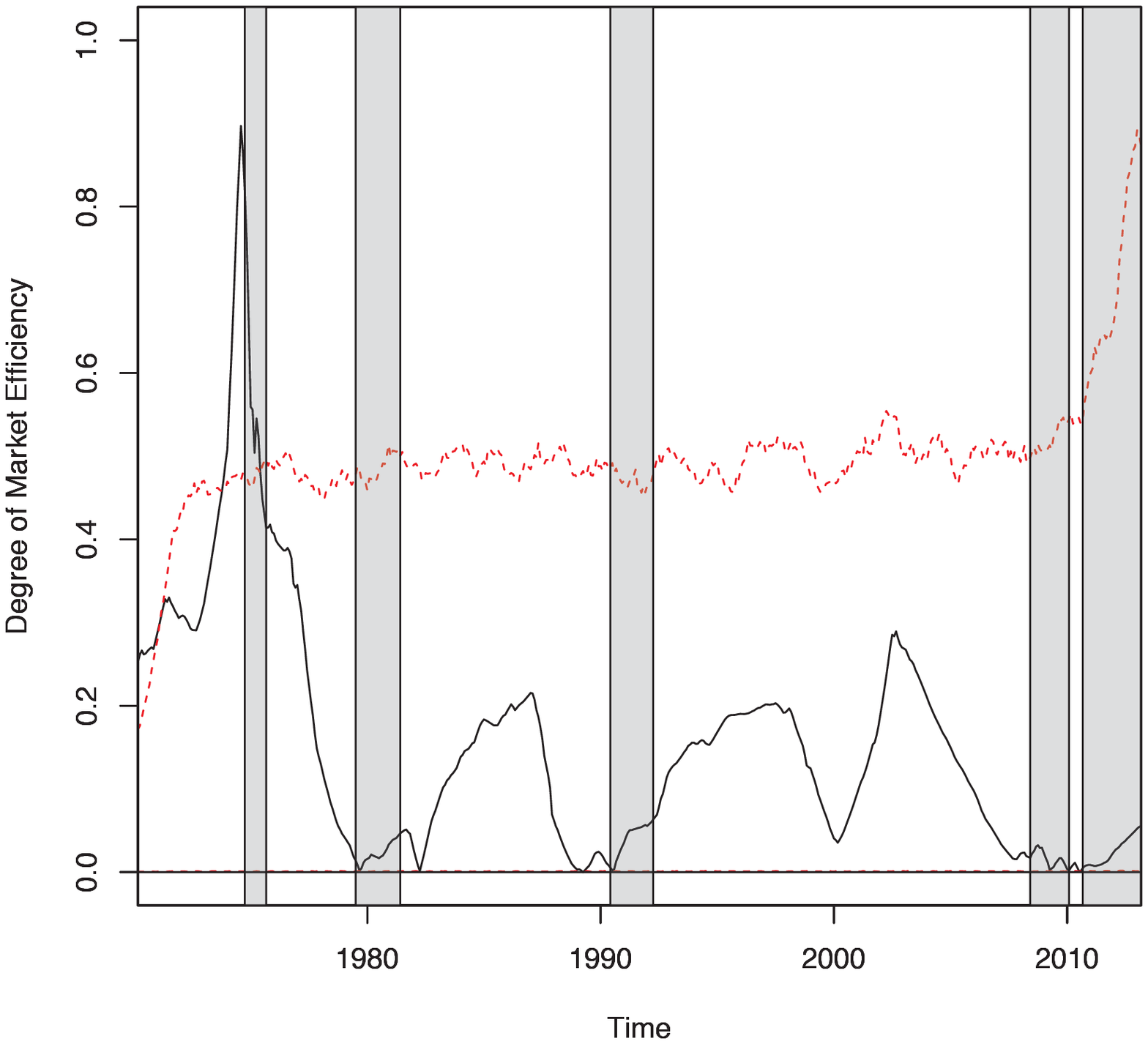}
\label{fig:figb13}}
\subfigure[Japan]{\includegraphics[width=.3\textwidth]{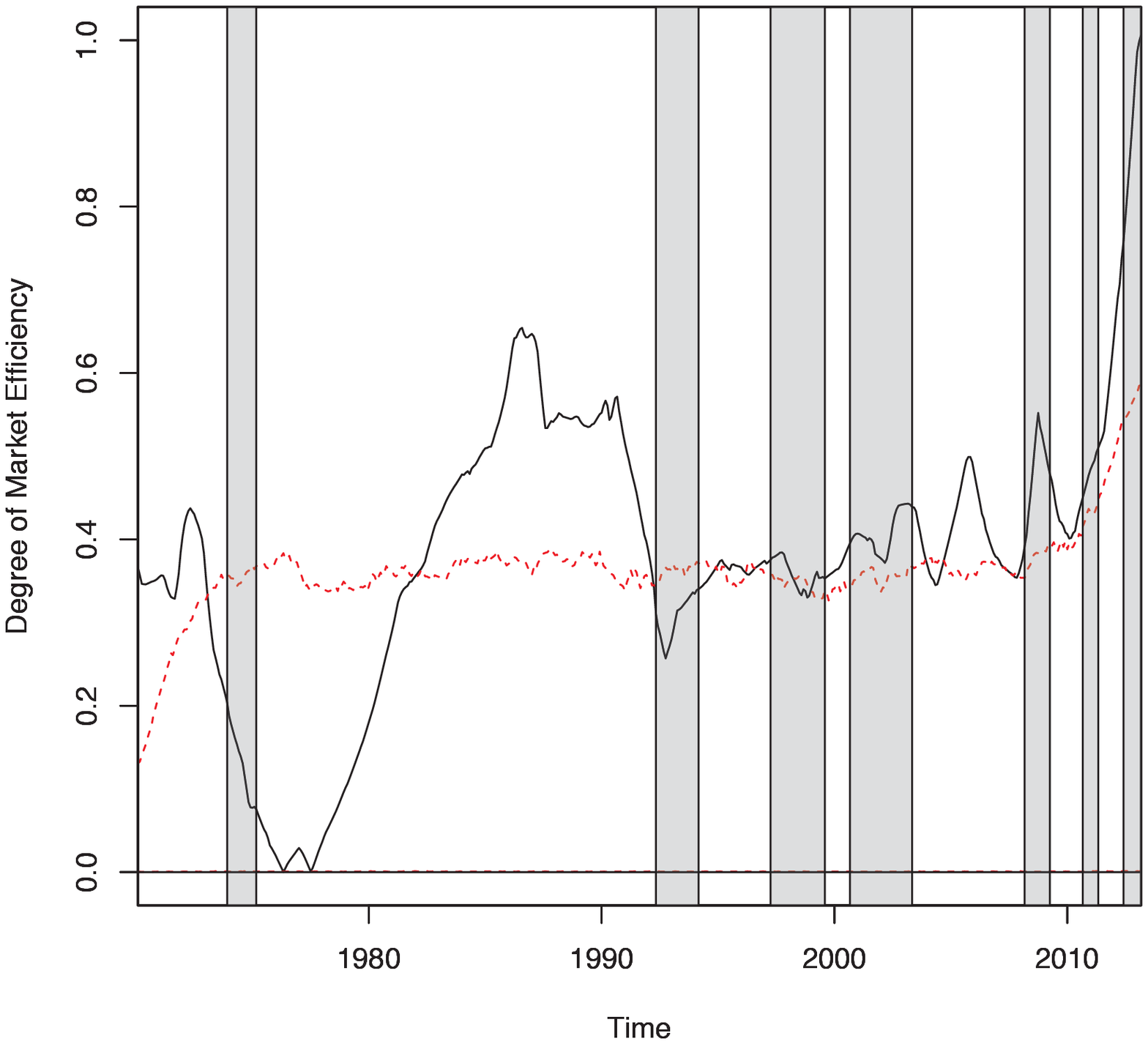}
\label{fig:figb21}}
\subfigure[Germany]{\includegraphics[width=.3\textwidth]{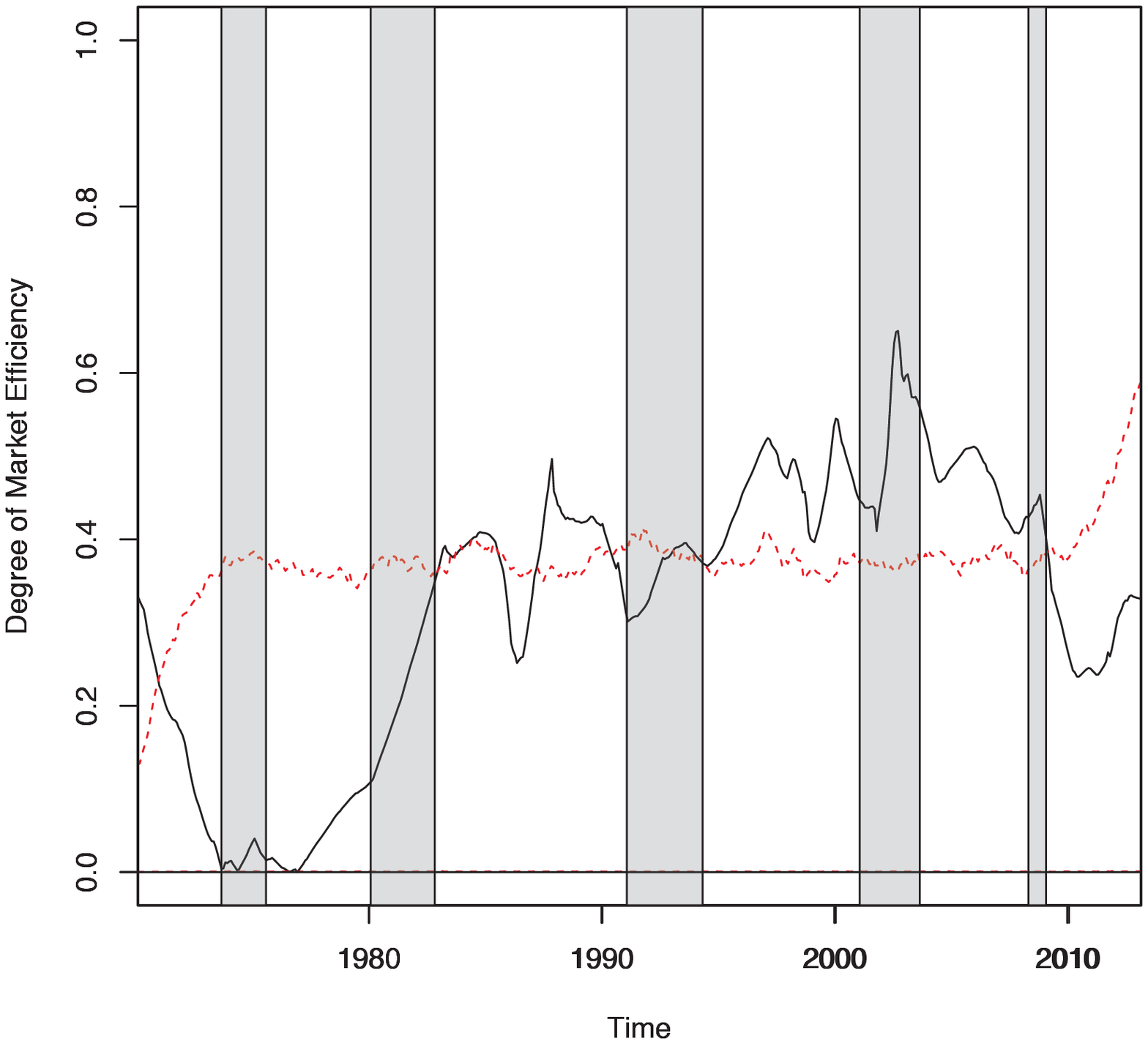}
\label{fig:figb22}}
\subfigure[France]{\includegraphics[width=.3\textwidth]{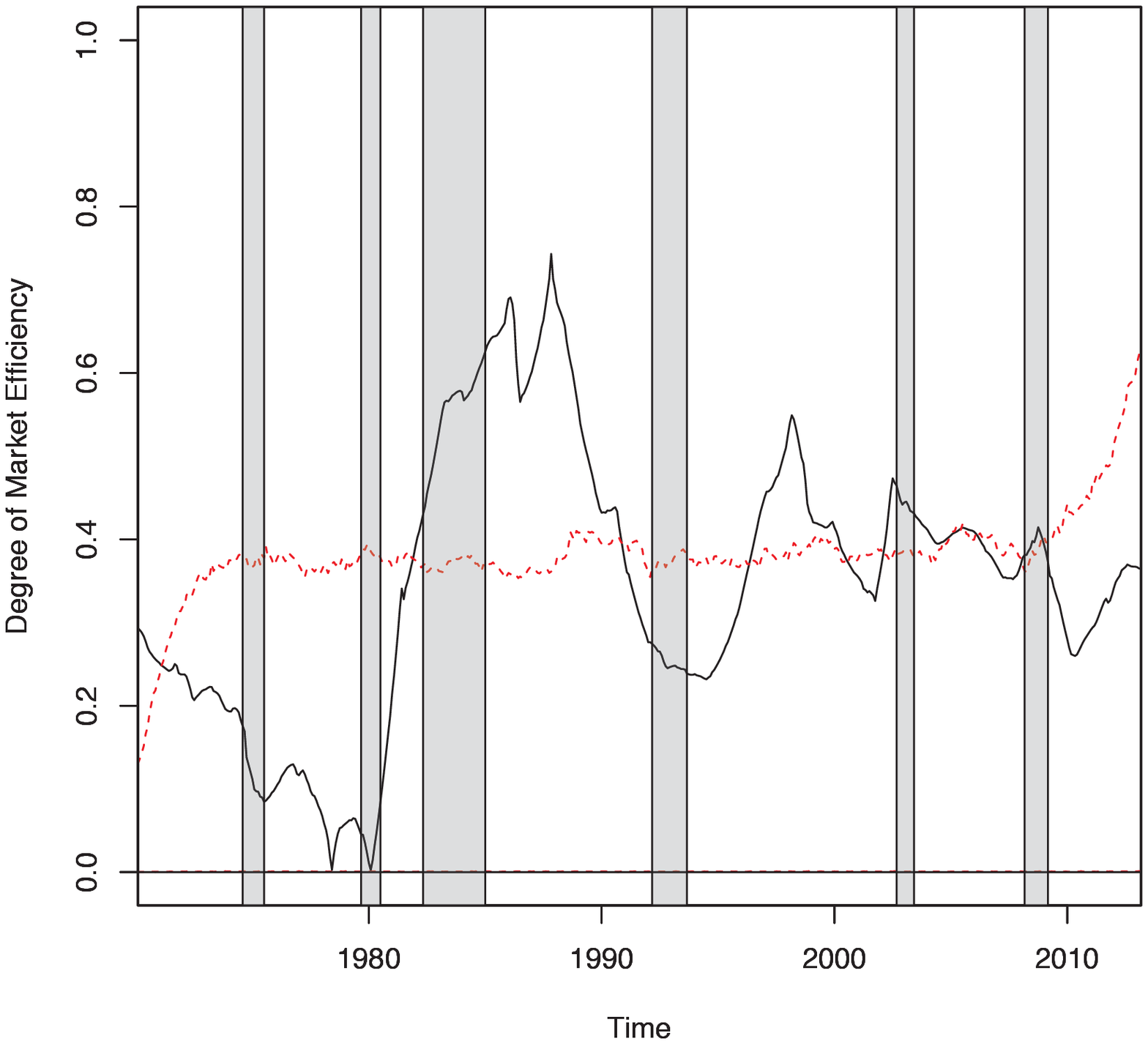}
\label{fig:figb23}}
\subfigure[Italy]{\includegraphics[width=.3\textwidth]{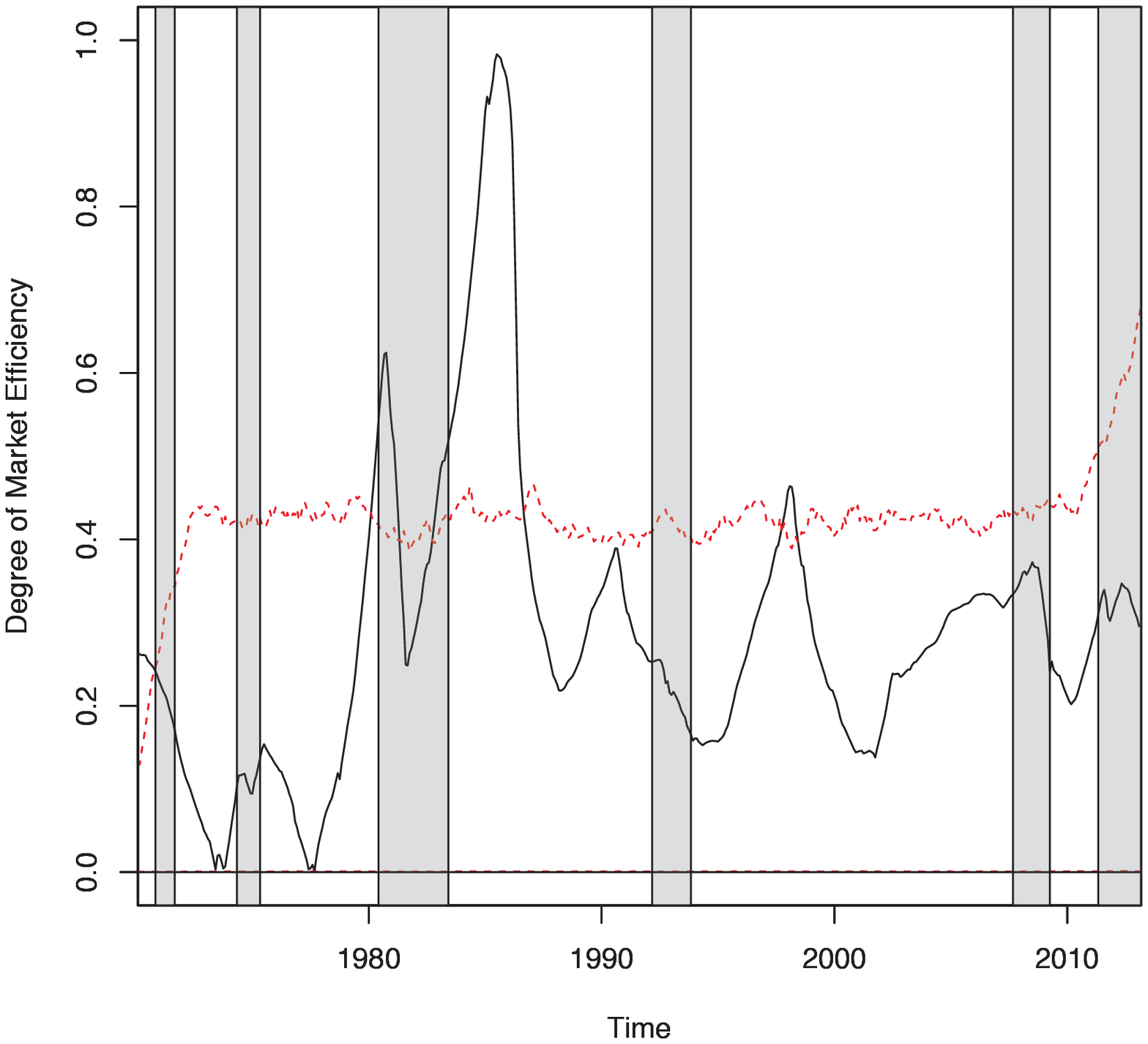}
\label{fig:figb31}}
\end{subfigmatrix}
\vspace*{3pt}
{\begin{center}
\begin{minipage}{420pt}
\footnotesize
\underline{Notes}:
\begin{itemize}
\item[(1)] The dashed red lines represent the 99\% confidence bands of the 
           time-varying spectral norm in case of efficient market. 
\item[(2)] We run 5000 times Monte Carlo sampling to calculate the confidence 
           bands.
\item[(3)] The shade areas represent recessions reported by the NBER business 
           cycle dates for the U.S. and the Economic Cycle Research Institute 
           business cycle peak and trough dates for the other countries.
\item[(4)] R version 3.1.0 was used to compute the estimates.
\end{itemize}
\end{minipage}
\end{center}}%

\end{figure}

Note that the degree of market efficiency is computed from the estimates of $\Phi_{t}\left(1\right)$, therefore, $\zeta_{t}$ is subject to sampling errors. Because of this, we provide the confidence band for $\zeta_{t}$ under the null hypothesis of market efficiency.\footnote{While we use the bootstrap method (i.e., using estimated residuals) to compute the confidence bands, the Monte Carlo method (i.e., using random draws from the normal distribution) can generate pretty much the same confidence bands. Therefore, our empirical results presented in this paper do not depend on which method is used to compute the statistical inference for the degree of market efficiency.} We do not find evidence of inefficient markets whenever estimated $\zeta_{t}$ is less than the upper limit of the confidence band; inefficient markets are detected with an estimated $\zeta_{t}$ that is larger than the upper limit. As detailed in Online Technical Appendix A.4, Monte Carlo simulations construct the confidence band by (i) generating multi-variate i.i.d. processes for $\ve y_{t}$, then, (ii) applying and estimating the TV-VAR model for those processes, and finally, (iii) computing $\zeta_{t}$.
\begin{center}
(Figures \ref{inter_market_fig1} to \ref{inter_market_fig5} around here)
\end{center}
Looking at North American markets, as demonstrated in Figure \ref{inter_market_fig1}, the markets become inefficient in the late 1980s and in the late 2000s; and almost completely inefficient in the early 2000s. Interestingly, these correspond well with Black Monday in 1987 and the Savings and Loan (S\&L) crisis in the late 1980s, the world financial crisis of 2008-2009, and the dot-com bubble burst in 2001. To provide a better understanding of the dynamics of the international market linkages, Figure \ref{inter_market_fig6} presents the degree of market efficiency for each of the seven countries. While the degrees for the U.S. and Canada have similar fluctuation patterns, inefficiency in the early 2000s seems to attribute to the Canadian market inefficiency.
\begin{center}
(Figure \ref{inter_market_fig6} around here)
\end{center}
The two largest stock markets become inefficient in the early to middle 1970s and the late 1980s. Yet, the U.K. market's inefficiency in the 1970s stands out. In addition to the period when the two largest markets are inefficient, the three largest stock markets, including Japan, become inefficient in the middle 2000s and after 2010.

Like other markets, the middle 1970s and 1980s are times for inefficient markets in Europe, the degree of market efficiency in the European markets is nevertheless increasing since the late 1990s. Why is the European markets' efficiency decreasing (the degree of market efficiency increasing) after the late 1990s? From individual degrees of efficiency presented in Figure \ref{inter_market_fig6}, it is hard to understand why. In addition, even when we compute the degree of market efficiency for each country individually (See Figure \ref{inter_market_fig6}), none of the European countries indicates such tendency. Remarkably, however, the introduction of the common currency euro in 1999 somewhat coincides the increase of the degree of market efficiency.

Finally, the efficiency of all seven countries changes over time. But importantly, in the early-to-middle 1970s, between the late 1980s and early 1990s, and after 2000 are three periods when the world (G7) markets indicate their inefficiency.

\subsection{Why Do Stock Markets Become Inefficient?}

There are, possibly, several reasons why stock markets become inefficient. First, as pointed out by \citet{ito2012eme} who focus solely on the US stock market, (i) people become irrational when they face extraordinary events such as severe recessions or financial crises\footnote{\citet{ito2012eme} utilize a much longer sample period for the US, and find that the world financial crisis in 2008 did not cause an inefficient market in the US. According to their study, inefficient markets emerged during: (i) the longest recession defined by NBER (1873-1879); (ii) the 1902-1904 recession; (iii) the New Deal era; and (iv) just after the very severe 1957-1958 recession.}; (ii) people are rational, but their stochastic discount factor changes for those time periods.\footnote{In such a case, the efficiency condition does not hold even with people's rational behavior.} It is also possible that the stock markets in our sample were indeed efficient, in combination with other stock markets that are not included in our sample. To fully understand why markets become inefficient, further investigation, perhaps with larger samples, is necessary.

\section{Concluding Remarks}\label{sec:inter_market_cr}

Stock market efficiency, especially the international linkage of stock markets, is studied. Paying attention to the stability of VAR coefficients, we determine that the time-varying (TV) VAR model fits better with the stock market data, with the help of \citetapos{hansen1992a} parameter constancy test. As an extension of \citet{ito2012eme}, we develop a new metric for market efficiency, namely the degree of market efficiency that is the spectral norm between the estimated time-varying parameter matrix and an identity matrix.

Our empirical results show that the degree of the market efficiency is, in fact, time-varying; and there are times when international markets are jointly efficient and inefficient. After considering five groups of countries, we interestingly find that (i) international stock markets are inefficient in the late 1980s; and (ii) European markets become jointly inefficient -- despite the fact that each country's stock market indicates otherwise -- after the late 1990s, which is almost corresponding to the commencement of their common currency.

However, caution should be taken in interpreting our results. For example, failing to find evidence of market efficiency in our sample does not necessarily mean that the markets are inefficient. This is because those markets are, possibly, jointly efficient with other markets that are not included in our sample. Also, further investigation should reveal the mechanism of how the degree of market efficiency fluctuates from time to time -- especially why the degree of market efficiency exhibits inefficient markets during extraordinary times -- by utilizing larger samples of data and by formal statistical tests.

\setcounter{table}{0}
\renewcommand{\thetable}{\arabic{table}}

\clearpage

\begin{table}[bp]
\caption{Descriptive Statistics and Unit Root Tests} \label{inter_market_table1}
\begin{center}
\begin{tabular}{ccccccccccccc} \hline\hline
& & Mean & SD & Min & Max & & ADF-GLS & Lags & $\hat\phi$ & &
 $\mathcal{N}$ \\\cline{3-6}\cline{8-10}\cline{12-12}
$R^{US}$ & & $0.0052$ & $0.0396$ & $-0.2324$ & $0.1596$ & & $-3.9408$ & $4$ & $0.5766$ & & $519$ & \\
$R^{CA}$ & & $0.0054$ & $0.0430$ & $-0.2365$ & $0.1496$ & & $-17.4826$ & $0$ & $0.2569$ & & $519$ & \\
$R^{GB}$ & & $0.0057$ & $0.0470$ & $-0.2468$ & $0.3685$ & & $-15.3813$ & $1$ & $0.2828$ & & $519$ & \\
$R^{JP}$ & & $0.0036$ & $0.0477$ & $-0.2515$ & $0.1597$ & & $-15.9591$ & $0$ & $0.3391$ & & $519$ & \\
$R^{DE}$ & & $0.0038$ & $0.0498$ & $-0.2924$ & $0.1386$ & & $-8.7929$ & $2$ & $0.4023$ & & $519$ & \\
$R^{FR}$ & & $0.0050$ & $0.0525$ & $-0.2004$ & $0.2040$ & & $-17.0041$ & $0$ & $0.2823$ & & $519$ & \\
$R^{IT}$ & & $0.0034$ & $0.0617$ & $-0.2290$ & $0.2061$ & & $-17.5358$ & $0$ & $0.2540$ & & $519$ & \\
\hline\hline
\end{tabular}
\vspace*{5pt}
{
\begin{minipage}{440pt}
\footnotesize
{\underline{Notes:}}
\begin{itemize}
\item[(1)] ``$R^{US}$'', ``$R^{CA}$'', ``$R^{GB}$'', ``$R^{JP}$'', 
           ``$R^{DE}$'', ``$R^{FR}$'', and ``$R^{IT}$'' denote the 
           returns of the Morgan Stanley capital index for the G7 countries 
           (United States, Canada, United Kingdom, Japan, Germany, France, 
           and Italy), respectively.
\item[(2)] ``ADF-GLS'' denotes the ADF-GLS test statistics, ``Lags'' 
           denotes the lag order selected by the MBIC, and ``$\hat\phi$'' 
           denotes the coefficients vector in the GLS detrended series 
           (see Equation (6) in \citet{ng2001lls}).
\item[(3)] In computing the ADF-GLS test, a model with a time trend 
           and a constant is assumed. The critical value at the 1\% 
           significance level for the ADF-GLS test is ``$-3.42$''.
\item[(4)] ``$\mathcal{N}$'' denotes the number of observations.
\item[(5)] R version 3.1.0 was used to compute the statistics.
\end{itemize}
\end{minipage}}%
\end{center}%
\end{table}%

\clearpage

\begin{landscape}
\begin{table}[bp]
\tiny
\caption{Standard VAR Estimations} \label{inter_market_table2}
\begin{center}
\begin{tabular}{clccccccccccccc} \hline\hline
  & & & $Constant$ & $R^{US}_{t-1}$ & $R^{CA}_{t-1}$ & $R^{GB}_{t-1}$ & $R^{JP}_{t-1}$ & $R^{DE}_{t-1}$ & $R^{FR}_{t-1}$ & $R^{IT}_{t-1}$ & & $\bar{R}^2$ & $L_C$ & \\\cline{4-11}\cline{13-14}
  & \multirow{2}*{(1) North America} & & & & & & & & & & & & & \\
  & & & & & & & & & & & & & & \\
  & \multicolumn{1}{r}{\multirow{2}*{$R^{US}_t$}} & & 0.0044 & 0.2230 & $-0.0306$ & \multirow{2}*{$-$} & \multirow{2}*{$-$} & \multirow{2}*{$-$} & \multirow{2}*{$-$} & \multirow{2}*{$-$} & & \multirow{2}*{0.0361} & \multirow{4}*{68.1322} & \\
  & & & [0.0018] & [0.0686] & [0.0732] & & & & & & & & & \\
  & \multicolumn{1}{r}{\multirow{2}*{$R^{CA}_t$}} & & 0.0038 & 0.1532 & 0.1277 & \multirow{2}*{$-$} & \multirow{2}*{$-$} & \multirow{2}*{$-$} & \multirow{2}*{$-$} & \multirow{2}*{$-$} & & \multirow{2}*{0.0603} & & \\
  & & & [0.0020] & [0.0904] & [0.0808] & & & & & & & & & \\\hline
  & \multirow{2}*{(2) U.S. and U.K.} & & & & & & & & & & & & & \\
  & & & & & & & & & & & & & & \\
  & \multicolumn{1}{r}{\multirow{2}*{$R^{US}_t$}} & & 0.0041 & 0.1274 & \multirow{2}*{$-$} & 0.1017 & \multirow{2}*{$-$} & \multirow{2}*{$-$} & \multirow{2}*{$-$} & \multirow{2}*{$-$} & & \multirow{2}*{0.0453} & \multirow{4}*{52.6841} & \\
  & & & [0.0019] & [0.0730] & & [0.0414] & & & & & & & & \\
  & \multicolumn{1}{r}{\multirow{2}*{$R^{GB}_t$}} & & 0.0035 & 0.2242 & \multirow{2}*{$-$} & 0.1726 & \multirow{2}*{$-$} & \multirow{2}*{$-$} & \multirow{2}*{$-$} & \multirow{2}*{$-$} & & \multirow{2}*{0.1000} & & \\
  & & & [0.0019] & [0.0772] & & [0.0636] & & & & & & & & \\\hline
  & \multirow{2}*{(3) (2) plus Japan} & & & & & & & & & & & & & \\
  & & & & & & & & & & & & & & \\
  & \multicolumn{1}{r}{\multirow{2}*{$R^{US}_t$}} & & 0.0041 & 0.1411 & \multirow{2}*{$-$} & 0.1103 & $-0.0343$ & \multirow{2}*{$-$} & \multirow{2}*{$-$} & \multirow{2}*{$-$} & & \multirow{2}*{0.0447} & \multirow{6}*{85.8165} & \\
  & & & [0.0019] & [0.0820] & & [0.0392] & [0.0481] & & & & & & & \\
  & \multicolumn{1}{r}{\multirow{2}*{$R^{GB}_t$}} & & 0.0035 & 0.2256 & \multirow{2}*{$-$} & 0.1735 & $-0.0036$ & \multirow{2}*{$-$} & \multirow{2}*{$-$} & \multirow{2}*{$-$} & & \multirow{2}*{0.0983} & & \\
  & & & [0.0019] & [0.0823] & & [0.0659] & [0.0490] & & & & & & & \\
  & \multicolumn{1}{r}{\multirow{2}*{$R^{JP}_t$}} & & 0.0020 & 0.1320 & \multirow{2}*{$-$} & 0.0373 & 0.2055 & \multirow{2}*{$-$} & \multirow{2}*{$-$} & \multirow{2}*{$-$} & & \multirow{2}*{0.0829} & & \\
  & & & [0.0021] & [0.0807] & & [0.0499] & [0.0590] & & & & & & & \\\hline
  & \multirow{2}*{(4) EU Countries} & & & & & & & & & & & & & \\
  & & & & & & & & & & & & & & \\
  & \multicolumn{1}{r}{\multirow{2}*{$R^{GB}_t$}} & & 0.0039 & \multirow{2}*{$-$} & \multirow{2}*{$-$} & 0.2475 & \multirow{2}*{$-$} & $-0.0603$ & 0.1476 & $-0.0347$ & & \multirow{2}*{0.0854} & \multirow{8}*{117.8807} & \\
  & & & [0.0018] & & & [0.0586] & & [0.0564] & [0.0833] & [0.0522] & & & & \\
  & \multicolumn{1}{r}{\multirow{2}*{$R^{DE}_t$}} & & 0.0026 & \multirow{2}*{$-$} & \multirow{2}*{$-$} & 0.0600 & \multirow{2}*{$-$} & 0.1692 & 0.0646 & 0.0097 & & \multirow{2}*{0.0630} & & \\
  & & & [0.0022] & & & [0.0485] & & [0.0596] & [0.0721] & [0.0444] & & & & \\
  & \multicolumn{1}{r}{\multirow{2}*{$R^{FR}_t$}} & & 0.0035 & \multirow{2}*{$-$} & \multirow{2}*{$-$} & 0.0747 & \multirow{2}*{$-$} & 0.0613 & 0.1507 & 0.0049 & & \multirow{2}*{0.0520} & & \\
  & & & [0.0023] & & & [0.0746] & & [0.0704] & [0.0764] & [0.0635] & & & & \\
  & \multicolumn{1}{r}{\multirow{2}*{$R^{IT}_t$}} & & 0.0021 & \multirow{2}*{$-$} & \multirow{2}*{$-$} & $-0.0590$ & \multirow{2}*{$-$} & $-0.0145$ & 0.2442 & 0.1221 & & \multirow{2}*{0.0619} & & \\
  & & & [0.0026] & & & [0.0649] & & [0.0895] & [0.1051] & [0.0742] & & & & \\\hline
  & \multirow{2}*{(5) G7 Countries} & & & & & & & & & & & & & \\
  & & & & & & & & & & & & & & \\
  & \multicolumn{1}{r}{\multirow{2}*{$R^{US}_t$}} & & 0.0042 & 0.01814 & $-0.0459$ & 0.1131 & $-0.0285$ & $-0.0336$ & 0.0064 & 0.0179 & & \multirow{2}*{0.0393} & \multirow{14}*{160.5454} & \\
  & & & [0.0019] & [0.0964] & [0.0793] & [0.0453] & [0.0502] & [0.0508] & [0.0597] & [0.0370] & & & & \\
  & \multicolumn{1}{r}{\multirow{2}*{$R^{CA}_t$}} & & 0.0039 & 0.1448 & 0.1241 & 0.0065 & 0.0653 & 0.0024 & $-0.0576$ & 0.0194 & & \multirow{2}*{0.0570} & & \\
  & & & [0.0020] & [0.0983] & [0.0819] & [0.0617] & [0.0531] & [0.0601] & [0.0669] & [0.0381] & & & & \\
  & \multicolumn{1}{r}{\multirow{2}*{$R^{GB}_t$}} & & 0.0034 & 0.2481 & $-0.0142$ & 0.1850 & 0.0110 & $-0.1094$ & 0.1027 & $-0.0390$ & & \multirow{2}*{0.1018} & & \\
  & & & [0.0018] & [0.1051] & [0.0891] & [0.0661] & [0.0515] & [0.0543] & [0.0729] & [0.0523] & & & & \\
  & \multicolumn{1}{r}{\multirow{2}*{$R^{JP}_t$}} & & 0.0020 & 0.1836 & $-0.0397$ & 0.0395 & 0.2175 & $-0.1273$ & 0.0397 & 0.0514 & & \multirow{2}*{0.0853} & & \\
  & & & [0.0021] & [0.0972] & [0.0771] & [0.0490] & [0.0615] & [0.0666] & [0.0642] & [0.0364] & & & & \\
  & \multicolumn{1}{r}{\multirow{2}*{$R^{DE}_t$}} & & 0.0022 & 0.2668 & $-0.0211$ & 0.0081 & $-0.0521$ & 0.1308 & 0.0248 & 0.0094 & & \multirow{2}*{0.0785} & & \\
  & & & [0.0022] & [0.0997] & [0.0812] & [0.0491] & [0.0496] & [0.0616] & [0.0604] & [0.0470] & & & & \\
  & \multicolumn{1}{r}{\multirow{2}*{$R^{FR}_t$}} & & 0.0033 & 0.2438 & $-0.1069$ & 0.0441 & $-0.0330$ & 0.0359 & 0.1333 & 0.0067 & & \multirow{2}*{0.0581} & & \\
  & & & [0.0024] & [0.1082] & [0.0955] & [0.0797] & [0.0578] & [0.0772] & [0.0782] & [0.0642] & & & & \\
  & \multicolumn{1}{r}{\multirow{2}*{$R^{IT}_t$}} & & 0.0017 & 0.2024 & $-0.0312$ & $-0.1029$ & $-0.0031$ & $-0.0491$ & 0.2136 & 0.1201 & & \multirow{2}*{0.0636} & & \\
  & & & [0.0026] & [0.1087] & [0.1107] & [0.0644] & [0.0602] & [0.0807] & [0.1103] & [0.0709] & & & & \\\hline\hline
\end{tabular}
\vspace*{2pt}
{
\begin{minipage}{480pt}
\tiny
{\underline{Notes:}}
\begin{itemize}
\item[(1)] ``$R^{US}_{t-1}$'', ``$R^{CA}_{t-1}$'', ``$R^{GB}_{t-1}$'', 
           ``$R^{JP}_{t-1}$'', ``$R^{DE}_{t-1}$'', ``$R^{FR}_{t-1}$'', 
           and ``$R^{IT}_{t-1}$'' denote the lagged returns of the Morgan 
           Stanley capital index for the G7 countries (United States, 
           Canada, United Kingdom, Japan, Germany, France, and Italy), 
           respectively.
\item[(2)] ``${\bar{R}}^2$'' denotes the adjusted $R^2$, and ``$L_C$'' 
           denotes \citetapos{hansen1992a} joint $L$ statistic with variance.
\item[(3)] \citetapos{newey1987sps} robust standard errors are in brackets.
\item[(4)] R version 3.1.0 was used to compute the estimates and the
	   statistics.
\end{itemize}
\end{minipage}}%
\end{center}%
\end{table}%
\end{landscape}

\clearpage

\clearpage

\clearpage

\clearpage

\clearpage

\clearpage

\end{document}